\documentclass[preprint]{elsarticle}

\usepackage[english]{babel}

\usepackage[normalem]{ulem} 
\usepackage[super]{nth} 
\usepackage{setspace}
\usepackage{amsmath, amsthm,amssymb}
\usepackage{float}
\usepackage{bm}
\usepackage{graphicx}

\renewcommand{\le}{\leqslant}
\renewcommand{\ge}{\geqslant}

\newtheorem{thm}{Theorem}

\begin{document}

\title{Predator-Prey Linear Coupling with Hybrid Species}

\author{Jean-Luc Boulnois}

\address{Babson College, Babson Park, Wellesley, Massachusetts 02457, jlboulnois@msn.com}

\begin{abstract}
The classical two-species non-linear Predator-Prey system, often used in population dynamics modeling, is expressed in terms of a single positive coupling parameter $\lambda$. Based on standard logarithmic transformations, we derive a novel $\lambda$-\textit{invariant} Hamiltonian resulting in two coupled first-order ODEs for ``hybrid-species'', \textit{albeit} with one being \textit{linear}; we thus derive a new exact, closed-form, single quadrature solution valid for any value of $\lambda$ and the system's energy. In the particular case $\lambda  = 1$ the ODE system completely uncouples and a new, exact, energy-only dependent simple quadrature solution is derived. In the case $\lambda \neq 1$ an accurate practical approximation uncoupling the non-linear system is proposed and solutions are provided in terms of explicit quadratures together with high energy asymptotic solutions. A novel, exact, closed-form expression of the system's oscillation period valid for any value of $\lambda$ and orbital energy is also derived; two fundamental properties of the period are established; for $\lambda = 1$ the period is expressed in terms of a universal  energy function and shown to be the shortest.  
\end{abstract}


\maketitle


\section{Introduction}

The historic Predator-Prey problem, also known as the Lotka-Volterra (``LV'') system of two coupled first-order nonlinear differential equations, has first been investigated in ecological and chemical systems \cite{volterra1926},\cite{lotka}. This classical problem models the competition of two isolated coexisting species: a `prey population' evolves while feeding from an infinitely large resource supply, whereas `predators' interact by exclusively feeding on preys, either through direct predation or as parasites. This idealized two-species model has further been generalized to interactions between multiple coexisting species in biological mathematics \cite{volterra1931}, ecology \cite{Chauvet}, virus propagation \cite {Chen-Charpentier}, and also in molecular vibration-vibration energy transfers \cite{treanor}. 
\newline

Let $u' \geqslant 0$ and $v' \geqslant 0$ be the respective instantaneous populations of preys and predators assumed to be continuous functions of time $t'$ : the net growth rates of each species is modeled as a system of two coupled first-order autonomous nonlinear ordinary differential equations (ODEs) according to
\begin{subequations}\label{EQ_1_}
    \begin{align}
        \frac{du'}{u'dt'} &=\alpha -\beta v' \quad     \text{for preys} \\
        \frac{dv'}{v'dt'} &=\gamma u'-\delta \quad     \text{ for predators}
    \end{align}
\end{subequations}

In the  classical LV model, $\alpha ,\beta ,\gamma ,\delta $ are assumed to be time-independent, positive, and constant: the rates $\alpha$ and $\delta$ represent self-interaction while the rates associated with $\beta$ and $\delta$ characterize inter-species interaction. In absence of predators, the natural exponential growth rate of the prey population is $\alpha$ ; when interacting with predators this population decreases at a rate modeled as $-\beta v'$. 
Similarly, when preys are scarce, the predator population decays at a rate $-\delta $ , and when feeding on preys its growth rate is modeled as  $\gamma u'$. 
\newline

Numerous solutions of the non-linear system \eqref{EQ_1_} using a variety of techniques have been proposed including trigonometric series \cite{frame}, Lambert W-functions \cite{shih1997}, \cite{shih2005}, mathematical transformations \cite{evans1999}, Taylor series expansions \cite{Scarpello}, perturbation techniques \cite{rao}, \cite{murty}, and numeric-analytic techniques \cite{chowdhury}. Also, an exact solution has been derived by Varma \cite{varma} in the special case when the rates $\alpha$ and $\delta$ are identical in magnitude, but with $\alpha =-\delta $, a condition which precludes population oscillation. The basic system \eqref{EQ_1_} is non-trivial and analytical closed-form solutions are unknown.

\section{Normalized Equations and Single Coupling Parameter}

Without any loss of generality, the system \eqref{EQ_1_} can further be simplified by simultaneously rescaling the predator and prey populations according to $v=\left(\beta /\alpha \right)v'$  and  $u=\left(\gamma /\delta \right)u'$ respectively, while also rescaling time through a ``\textit{stretched}'' time without unit  $t=\sqrt{\alpha \delta } t'$. Upon introducing the positive coupling parameter $\lambda $, ratio of the respective growth and decay rates of each species taken separately, defined as
\begin{equation} \label{EQ_2_} 
\lambda =\sqrt{\frac{\alpha }{\delta } }  
\end{equation}

a normalized form of the LV system is obtained as a set of two coupled nonlinear first-order ODEs exclusively depending on this single coupling ratio  $\lambda $  according to
\begin{subequations}\label{EQ_4_}
    \begin{align}
        \dot{u}&=\lambda u\left(1-v\right) \quad     \text{for preys} \\
        \dot{v}&=\frac{1}{\lambda } v\left(u-1\right) \quad     \text{ for predators}
    \end{align}
\end{subequations}

Here the ``dot'' on  $\dot{u}$  and  $\dot{v}$  indicates a derivative with respect to the  time  $t$: in the sudden absence of coupling between species ($\beta = \gamma = 0$), the prey population would grow at an exponential rate $\lambda $ while predators would similarly decay at an inverse rate  $-1/\lambda $ from their respective positive initial values. Remarkably, the normalized ODE system \eqref{EQ_4_}  is invariant in the transformation  $u\to v$  together with  $\lambda \to -1/\lambda $: this fundamental property, subsequently referred to as ``$\lambda $-\textit{invariance}", is extensively used throughout to considerably simplify the LV problem analysis.

Since the original  publications \cite{volterra1926}, \cite{lotka}, the system \eqref{EQ_4_} has been known to possess a dynamical invariant or ``constant of motion  $K$'' expressed here in $\lambda$-\textit{invariant} form 
\begin{equation} \label{EQ_5_} 
\frac{1}{\lambda } u+\lambda v-\ln\left(u^{\frac{1}{\lambda } } v^{\lambda } \right)=K 
\end{equation}

In the following sections, through a particular Hamiltonian transformation combined with a suitable linear change of variables we introduce a novel  \textit{$\lambda$-invariant}  Hamiltonian based on new ``hybrid-species'' that reduces the system \eqref{EQ_4_} to a new set of two coupled first-order ODEs with one being \textit{linear}. Upon exploiting this linearity, a new, exact analytical solution is derived for one hybrid-species in terms of a simple quadrature: we then proceed with an original method to uncouple the system and derive complete, closed-form quadrature solutions of the LV problem. The population oscillation period is further derived in terms of a unique energy function and two fundamental properties are established.

\section{Solutions with Hybrid Predator-Prey Species}

The logarithmic functional transformation originally introduced by Kerner \cite{kerner1964} reduces the normalized LV system \eqref{EQ_4_} to a Hamiltonian form: the coupling between the respective species is modified through a change of variables according to
\begin{equation}
y=ln(u) \text{ and } x=ln(v) \text{ with } y\in (-\infty , +\infty ), x\in (-\infty , +\infty )      \label{EQ_6_}
\end{equation} 

The LV system \eqref{EQ_4_} for the respective ``logarithmic'' prey-like and predator-like species $y(t)$ and $x(t)$  becomes
\begin{equation} \label{EQ_7_} 
\begin{array}{l} {\dot{y}=\lambda (1-e^{x} )} \\ {\dot{x}=\frac{1}{\lambda } (e^{y} -1)} \end{array} 
\end{equation}

Similarly to Eq. \eqref{EQ_5_} this $\lambda$-\textit{ invariant} system \eqref{EQ_7_} admits a primary conservation integral $H$ expressed as the linear combination of two positive convex functions
\begin{equation} \label{EQ_8_} 
H(x,y)=\lambda (e^{x} -x-1)+\frac{1}{\lambda } (e^{y} -y-1) 
\end{equation} 

As already established \cite{plank}, \cite{kerner1997}, $H(x,y)$  is the Hamiltonian of the conservative LV system since Eqs. \eqref{EQ_7_} satisfy Hamilton's equations with $x$  as the coordinate conjugate to the canonical momentum $y$. Equation \eqref{EQ_8_} expresses the conservative coupling between species $x(t)$ and $y(t)$: it is further rendered $\lambda $-\textit{ invariant}  by introducing a scaled Hamiltonian $h(x,y)$ with total constant positive energy simply labeled $h$, according to
\begin{equation} \label{EQ_9_} 
H\left(x,y\right)=\left(\lambda +\frac{1}{\lambda } \right)h\left(x,y\right) 
\end{equation}

We introduce a $\lambda $-\textit{ invariant}  linear first-order ODE between the species  $x(t)$ and $y(t)$  by further combining the system \eqref{EQ_7_} with \eqref{EQ_8_} and \eqref{EQ_9_}
\begin{equation} \label{EQ_10_} 
\dot{x}-\dot{y}-\left(\lambda x+\frac{y}{\lambda } \right)=\left(\lambda +\frac{1}{\lambda } \right)h 
\end{equation}

Equation \eqref{EQ_10_} suggests introducing a \textit{$\lambda $-invariant}  linear transformation of the set $\{$\textit{x(t), y(t)}$\}$ to a new set $\{$\textit{$\xi$(t)}, \textit{$\eta$(t)}$\}$ representing the symbiotic coupling between "\textit{hybrid} predator-prey species" 
\begin{subequations} \label{EQ_11_} 
     \begin{align}
     \xi&= \frac{\lambda x + \frac{1}{\lambda} y}{\lambda+\frac{1}{\lambda}} \label{EQ_11_a}\\
     \eta&= \frac{x-y}{\lambda+\frac{1}{\lambda}}  \label{EQ_11_b}
     \end{align} 
\end{subequations}

The original Hamiltonian \eqref{EQ_8_} together with \eqref{EQ_9_} and the linear transformation \eqref{EQ_11_} then becomes
\begin{equation} \label{EQ_12_} 
h(\eta ,\xi )=\frac{\lambda e^{\frac{\eta }{\lambda } } +\frac{1}{\lambda } e^{-\lambda \eta } }{\lambda +\frac{1}{\lambda } } e^{\xi } -\xi -1 
\end{equation}

Here $h\left(\eta ,\xi \right)$ is a new Hamiltonian for the coordinate $\eta$ and conjugate momentum $\xi$. Notice that for small amplitudes, $h\left(\eta ,\xi \right)$ is the Hamiltonian of a harmonic oscillator.
Upon further introducing the following $\lambda$-\textit{invariant} $G$-function 
\begin{equation}  \label{EQ_13_}
G_{\lambda } (\eta )=\frac{\lambda e^{\frac{\eta }{\lambda } } +\frac{1}{\lambda } e^{-\lambda \eta } }{\lambda +\frac{1}{\lambda } }    \quad \text{ with }   G_{\lambda } (\eta)=G_{1/\lambda } (-\eta )   \quad  (\lambda \textit{-invariance})    
\end{equation}

the conservation relationship \eqref{EQ_12_} between the conjugate functions  \textit{$\eta$(t)}  and \textit{ $\xi$(t)}  is recast into a compact form which provides a natural separation of variables
\begin{equation} \label{EQ_14_} 
G_{\lambda } (\eta )=(h+1+\xi )e^{-\xi }  
\end{equation}

In the following we define the function  \textit{U($\xi$)}  that appears throughout as  
\begin{equation} \label{EQ_15_} 
U(\xi )=(h+1+\xi )e^{-\xi }  
\end{equation}

Even though still nonlinear, the fundamental conservation relationship \eqref{EQ_14_} partially uncouples the  \textit{$\xi$(t)-}function from the  \textit{$\eta$(t)-}function, resulting in three essential \textit{G}-function properties:

\begin{enumerate}
\item  the system's energy $h\geqslant  0$ is explicitly associated with the function $U(\xi )$ only; 

\item  the positive function $G_{\lambda } (\eta )$ is a generalized hyperbolic cosine function that reaches its minimum $G_{\lambda}= 1$  at $\eta = 0$  for any value of $\lambda $ : hence its inverse function  $G_{\lambda }^{-1} $  exists, and, for any value of  $\lambda $ , Eq. \eqref{EQ_14_} admits two respective positive and negative roots  $\eta ^{\pm } (\xi ,\lambda )$ functions of $\xi$ only satisfying
\begin{equation} \label{EQ_16_} 
\eta ^{\pm } (\xi ,\lambda )=G_{\lambda }^{-1} \big(U(\xi )\big) 
\end{equation}

\item  since the $\eta$-function is associated with the coupling ratio $\lambda $ only, $\lambda $\textit{-invariance} of the $G$-function \eqref{EQ_13_}  implies that, for a given $\lambda$, any positive solution  $\eta ^{+} (\xi ,\lambda )$  is directly derived from the negative solution associated with the ratio $1/\lambda$, and reciprocally 
\begin{equation} \label{EQ_17_} 
\eta ^{\pm } (\xi ,\lambda )=-\eta ^{\mp } (\xi ,1/\lambda ) 
\end{equation} 
\end{enumerate}

From Eq. \eqref{EQ_15_} the hybrid-species population $\xi (t)$  thus oscillates between the $\lambda $-independent respective negative and positive roots $\xi ^{-} (h)$ and  $\xi ^{+} (h)$, solutions of the equation $U(\xi )=1$, solely dependent on the system's energy $h$ as displayed in Table \ref{table:1} for several increasing values of $h$
\begin{equation} \label{EQ_18_} 
e^{\xi } -\xi -1=h  \quad \text{ with } h\geqslant  0
\end{equation}

\begin{table}[H]
{\small

\begin{tabular}{|p{0.38in}|p{0.38in}|p{0.38in}|p{0.38in}|p{0.38in}|p{0.38in}|p{0.38in}|p{0.38in}|p{0.38in}|} \hline 
\textit{h} & 0.3 & 0.5 & 1 & 2 & 3 & 5 & 7 & 10 \\ \hline 
$\xi ^{-} (h)$ & -0.889 & -1.198 & -1.841 & -2.948 & -3.981 & -5.998 & -8.000 & -11.00 \\ \hline 
$\xi ^{+} (h)$ & 0.686 & 0.858 & 1.146 & 1.505 & 1.749 & 2.091 & 2.336 & 2.611 \\ \hline 
\end{tabular}
}
\caption{Roots of  $e^{\xi } -\xi -1=h$  as a function of the energy $h$   from Eq. \eqref{EQ_18_}}
\label{table:1}
\end{table}

In the $\xi -\eta $ plane, Eq. \eqref{EQ_14_} represents a closed-orbit mapping around the fixed point $(0,0)$.  On the $\eta =0$ horizontal axis this orbit is bounded by the limits $\xi ^{-} (h)$ and  $\xi ^{+} (h)$, and since  \textit{U($\xi$)}  admits a maximum $e^{h} $ located at $\xi = -h$ , it is also bounded by the two respective positive and negative roots solutions of the equation  $\eta ^{\pm } (-h,\lambda )=G_{\lambda }^{-1} (e^{h} )$. 
For any given energy  $h$  this orbit consists of two respective branches  $\eta ^{+} (\xi ,\lambda )$ and  $\eta ^{-} (\xi ,\lambda )$ as displayed on Fig. 1 where the respective values chosen are $h = 2$ and coupling ratios  $\lambda = 2$  and   $\lambda = 1/2$. Per Eq. \eqref{EQ_17_}, the respective branches associated with the  $\lambda$  and  $1/\lambda$-mappings are readily observed to be symmetric with respect to the  $\eta = 0$  axis.
\newline

Except when $\lambda = 1$, algebraic solutions of Eq.\eqref{EQ_16_} may generally not be obtained directly.  However, for any value $\xi \in \{{\xi }^-(h),\ {\xi }^+(h)\}$  the two roots $\eta ^{\pm } (\xi ,\lambda )$ of  Eq. \eqref{EQ_16_} may numerically be obtained  through a standard "Newton-Raphson" algorithm.  Appendix 1 establishes that each root admits lower and upper bounds for any value of  \textit{U($\xi$)}, thereby ensuring  algorithm convergence.

\begin{figure}[H]
  \centering
  \includegraphics[width=11cm]{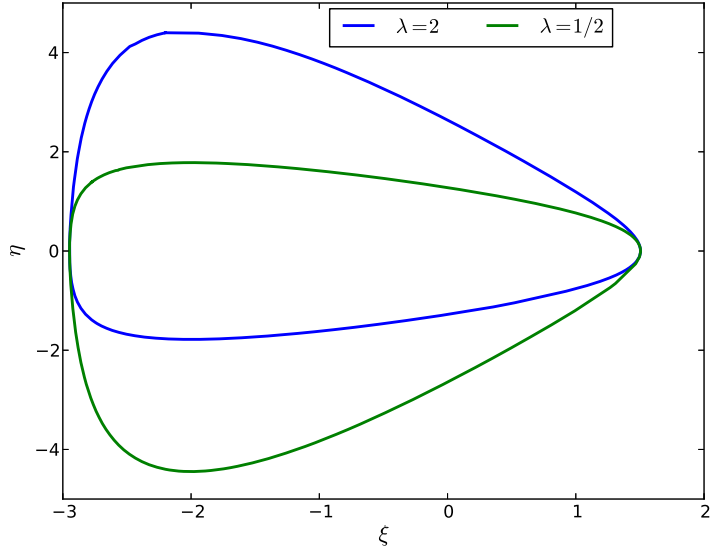}
  \caption{ $\xi -\eta $  Mapping for $\lambda = 2$ and  $\lambda = 1/2$, and energy $h = 2$}
  \label{fig:Fig1}
\end{figure}

Lastly, upon inserting the linear transformation \eqref{EQ_11_} into the modified LV system \eqref{EQ_7_}, or equivalently using the standard Hamilton equations with  Eq. \eqref{EQ_12_}, a new semi-linear system of coupled  \nth{1} order ODEs is obtained 
\begin{subequations}\label{EQ_19_}
    \begin{align}
        \dot{\eta }&=\xi + h \label{EQ_19_a}\\
        \dot{\xi }&=-G'_{\lambda } (\eta )e^{\xi } \label{EQ_19_b}
    \end{align}
\end{subequations}

The solution of the system \eqref{EQ_19_}, in which $G'_{\lambda } (\eta )$ is the derivative $G'_{\lambda } (\eta )=dG_{\lambda } /d\eta $, represents the  time-evolution of the hybrid-species $\eta (t)$  and $\xi (t)$, \textit{albeit} due to the linear transformation \eqref{EQ_11_}, the first coupled equation \eqref{EQ_19_a} becomes \textit{linear} since it directly expresses ODE \eqref{EQ_10_}. Remarkably, as a result of this hybrid-species transformation, up to the constant energy $h$, the time derivative of the function  $\eta (t)$ is directly equal to the instantaneous value of the species population $\xi (t)$, considerably simplifying the solution of \eqref{EQ_19_}.
The exact solution of the LV problem is then derived by integration of the linear ODE \eqref{EQ_19_a} as a \textit{simple closed-form quadrature} for  $t(\xi)$, time as a function of $\xi$: upon using the initial conditions $\eta _{0} =0$ and $\xi _{0} =\xi ^{\pm } (h)$ when  $t = 0$, the exact LV solution  corresponding to the respective negative and positive branches $\eta ^{-} (\xi ,\lambda )$ and $\eta ^{+} (\xi ,\lambda )$ simply becomes

\begin{equation} \label{EQ_20_} 
t(\xi )=\int _{\xi ^{\pm } }^{\xi }\frac{d\eta ^{\pm } (x,\lambda )}{h+x}   
\end{equation} 

This quadrature is not divergent at  $x=-h$ , since the differential $d\eta $ in Eq. \eqref{EQ_16_} contains the derivative  $U'(\xi )=-(h+\xi )e^{-\xi } $ in the numerator. Upon using the same initial conditions for  $\eta _{0} $ and  $\xi _{0} $, the solution \eqref{EQ_20_} is expressed in terms of the function  $\eta ^{\pm } (\xi ,\lambda )$ itself through a standard integration by parts  in which the singularity at $\xi = -h$  is further eliminated by adding and subtracting the expression  $\frac{\eta ^{\pm } (-h,\lambda )}{h+\xi } $  in the integral. The final, exact, closed-form, regular solution of the entire LV problem for any value of the coupling ratio $\lambda$ and any value of the orbital energy $h$ is thus explicitly expressed as a \textit{simple quadrature}  over each of the two branches $\eta ^{\pm } (\xi ,\lambda )$ solutions of  \eqref{EQ_16_}
\begin{equation} \label{EQ_21_} 
t(\xi )=\frac{\eta ^{\pm } (\xi ,\lambda )-\eta ^{\pm } (-h,\lambda )}{h+\xi } +\frac{\eta ^{\pm } (-h,\lambda )}{h+\xi ^{\pm } } +\int _{\xi ^{\pm } }^{\xi }\frac{\eta ^{\pm } (x,\lambda )-\eta ^{\pm } (-h,\lambda )}{(h+x)^{2} }  dx 
\end{equation} 

This exact solution is further analyzed in the following section. Numerical solutions for $\xi (t)$ and $\eta (t)$ are also obtained by integrating Eqs. \eqref{EQ_19_} using a standard fourth-order Runge-Kutta (RK4) method as presented in Fig. 2  for values of $h$  and $\lambda$  exactly identical to those of Fig. 1, together with  initial conditions $\eta _{0} $ and $\xi _{0} $ defined above. 
The function $\xi (t)$ is observed to principally depend on two time constants: a quasi-exponential increase at a rate of order $\lambda$ followed by an exponential decrease at a rate $-1/\lambda$. As expected from $\lambda$-\textit{invariance}  \eqref{EQ_17_} the two functions $\xi (t)$ respectively corresponding to the coupling ratio $\lambda = 2$ and its inverse $\lambda = 1/2$ are mirrors of each other; so are the functions $\eta (t)$, but with the change $\eta \to -\eta $.
\newline

It may generally not be possible to algebraically solve \eqref{EQ_16_} for $\eta (\xi ,\lambda )$ for insertion into the exact solution \eqref{EQ_21_}. 
A strategy consists in eliminating the $\eta$-dependence in \eqref{EQ_19_b} and seeking an ODE for $\xi(t)$ only: upon explicitly relating $G_{\lambda } (\eta )$ to its derivative ${G'}_{\lambda } (\eta )$ and expressing the latter as an analytical function of $\xi$ only through \eqref{EQ_14_}, a critical relationship is derived below.

\begin{figure}[H]

  \centering
  \includegraphics[width=11cm]{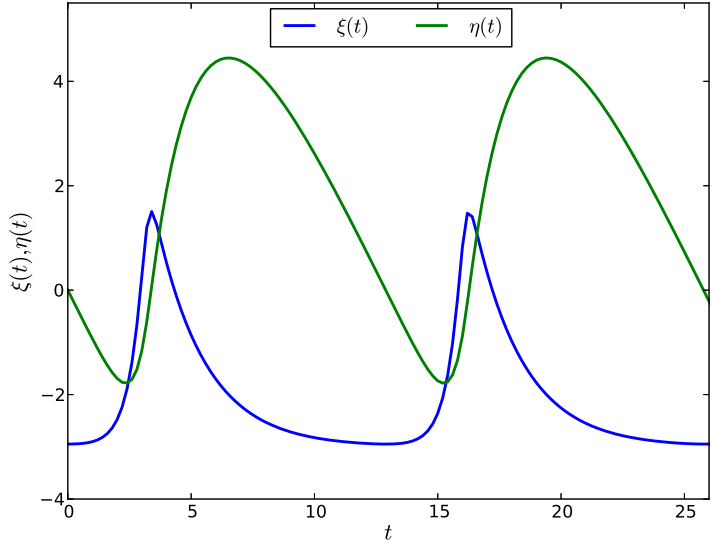}
  \caption{Solutions for $\xi (t)$ and  $\eta (t)$ as a function of time $t$  with $\lambda = 2$ and energy $h = 2$: numerical integration of Eq. \eqref{EQ_19_} by RK4}
  \label{fig:Fig2}
\end{figure}

\subsection*{Case $\lambda = 1$} 

The particular $\lambda = 1$ case is  exactly solved  since an explicit relationship exists between $G_{\lambda}$  and ${G'}_{\lambda}$ : it enables to entirely uncouple the ODE system \eqref{EQ_19_} and provides exact closed-form solutions for  $\xi(t)$ and $\eta(t)$ in terms of simple quadratures.
\newline

In this case, the $G$-function \eqref{EQ_13_} (omitting the index for simplicity) reduces to the hyperbolic cosine function ; the conservation equation \eqref{EQ_14_}  becomes
\begin{equation} \label{EQ_22_} 
G(\eta )=\cosh (\eta )=(h+1+\xi )e^{-\xi }  
\end{equation} 

The resulting  $\xi$ -- $\eta$   closed-orbit mapping is symmetric: on the $\xi$-axis, for any value of the orbital energy $h$,  the mapping is bounded by $\xi ^{-} (h)$ and $\xi ^{+} (h)$ defined in \eqref{EQ_18_}; the two symmetric branches  $\eta ^{\pm } (\xi )$  are explicitly expressed in terms of the inverse hyperbolic cosine function
\begin{equation} \label{EQ_23_} 
\eta ^{\pm } (\xi )=\pm \cosh ^{-1} \bigl((h+1+\xi )e^{-\xi } \bigr)
\end{equation}

Equation \eqref{EQ_23_} again establishes the symbiotic coupling between the hybrid species $\eta$ and $\xi$.  In this $\lambda = 1$ case, the explicit relationship sought earlier in the discussion of  \eqref{EQ_19_b} between $G(\eta )$ and its derivative $G'(\eta ) = \sinh(\eta )$ is
\begin{equation} \label{EQ_24_} 
G'(\eta )=\pm (G^{2} -1)^{1/2}  
\end{equation} 

Upon inserting \eqref{EQ_24_} together with \eqref{EQ_22_} into \eqref{EQ_19_b}, the nonlinear LV problem completely uncouples, consisting in the \nth{1} order \textit{linear} ODE  \eqref{EQ_19_a} together with a  \nth{1} order nonlinear autonomous ODE for the species  $\xi$ population
\begin{subequations}\label{EQ_25_}
    \begin{align}
       \dot{\eta }&=\xi +h \label{EQ_25_a}\\
        \dot{\xi }&=\pm e^{\xi } \bigl( (U(\xi ))^{2} -1 \bigr)
^{1/2} =\pm \bigl( (h+1+\xi )^{2} -e^{2\xi } \bigr)
^{1/2} \label{EQ_25_b}
    \end{align}
\end{subequations}

The linear equation \eqref{EQ_25_a} is directly solved by inserting $\eta (\xi)$ from \eqref{EQ_23_} into the solution \eqref{EQ_21_}. Together with  $U(\xi )$  defined in \eqref{EQ_15_}, the exact, closed-form analytic solution on the interval $\xi ^{-} \le \xi \le \xi ^{+} $ is thus expressed as a simple \textit{quadrature} in terms of elementary functions  
\begin{multline}
 \label{EQ_26_} 
t(\xi )=\frac{\cosh ^{-1} (e^{h} )-\cosh ^{-1} \bigl(U(\xi )\bigr)}{h+\xi } -\frac{\cosh ^{-1} (e^{h} )}{h+\xi ^{-} } + \\
\int _{\xi ^{-} }^{\xi }\frac{\cosh ^{-1} (e^{h} )-\cosh ^{-1} \bigl( U(x) \bigr) }{(h+x)^{2} }  dx 
\end{multline}

By applying l'H\^{o}pital's rule, it is readily verified that the integrand in \eqref{EQ_26_} is regular at $\xi = -h$. 
Figure 3 presents the $\xi (t)$-solution obtained by numerical integration of  \eqref{EQ_26_} for an energy $h = 2$. 
The complete solution of the  LV problem for $\lambda = 1$ is finalized for $\eta (t)$ by inserting $\xi (t)$ derived above into Eq. \eqref{EQ_23_}.

\begin{figure}[H]
  \centering
  \includegraphics[width=11cm]{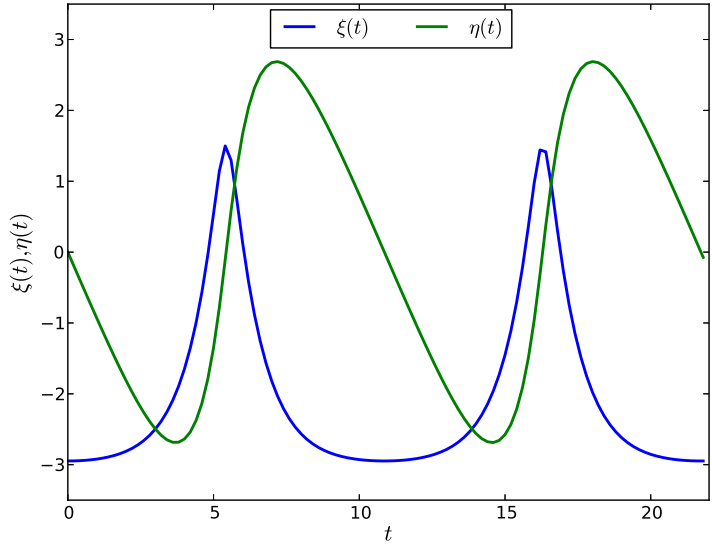}
  \caption{Solutions for $\xi (t)$ and  $\eta (t)$ as a function of time $t$  obtained by numerical integration of the quadrature solution Eq. \eqref{EQ_26_} with $\lambda = 1$  and energy $h = 2$}
  \label{fig:Fig4}
\end{figure}

Another expression for  $t(\xi )$  may be obtained by integrating $\xi (t)$ over the positive root in \eqref{EQ_25_b}, yielding a  simple alternative \textit{quadrature} solution 
\begin{equation} \label{EQ_27_} 
t(\xi )=\int _{\xi ^{-} }^{\xi }\frac{dx}{\sqrt{(h+1+x)^{2} - e^{2x} } }  
\end{equation}

It is readily verified that upon inserting $U(x)$  into the integrand of  \eqref{EQ_27_} and integrating by parts the resulting expression is identical to that of solution \eqref{EQ_26_}. The integrand of  \eqref{EQ_27_} has a weak singularity of the square root type at the respective limits $\xi ^{-} (h)$ and $\xi ^{+} (h)$, but is strictly continuous  and the integral is absolutely convergent. Finally, even though the oscillation of the hybrid-species population $\xi (t)$ is not expressed as an explicit function of time $t$, the function  \textit{t($\xi$)}  being monotonic and continuous on each integration interval for $\xi$, its inverse function $\xi (t)$, which uniquely depends on the energy level $h$, exists and is monotonic and continuous on each interval. The exact solution \eqref{EQ_27_} is similar in form to a solution derived by Evans and Findley (Eq. (17) in \cite{evans1999}); however, this integral expression lends itself to simpler analytical or numerical integration by standard methods. An exact expression for \eqref{EQ_27_} is further derived in Appendix 2 in terms of a series of exponential integrals. 




\subsection*{Case  $\lambda \ne 1$}

In the general case when $\lambda \ne 1$ the relationship between $G_{\lambda } (\eta )$ and its derivative $G'_{\lambda } (\eta )$ is obtained by observing that
\begin{equation} \label{EQ_28_} 
G'_{\lambda } (\eta )=\frac{e^{\frac{\eta }{\lambda } } -e^{-\lambda \eta } }{\lambda +\frac{1}{\lambda } } \quad  \text{ with }  G'_{\lambda } (\eta )=-G'_{1/\lambda } (-\eta ) \quad     (\lambda\textit{-invariance}) 
\end{equation}

Upon eliminating $\eta$ between Eqs. \eqref{EQ_13_} and \eqref{EQ_28_}, an implicit non-linear \nth{1} order ODE relating $G$ to its derivative $G'$ is derived (for clarity the index $\lambda$ is omitted in the remainder of this section) 
\begin{equation} \label{EQ_29_} 
\left( G+\frac{1}{\lambda } G'\right)^{\lambda } (G-\lambda G')^{1/\lambda } =1 
\end{equation}

Equation \eqref{EQ_29_} is completely invariant in the change $\lambda \to -1/\lambda $, or equivalently changing $\lambda \to 1/\lambda $ together with  $G'\to -G'$. 
As a result, similar to Eq. \eqref{EQ_24_}, in the $G-G'$ phase space, Eq. \eqref{EQ_29_} represents the positive and negative branches of a ``skewed'' hyperbola with orthogonal asymptotes, respectively $G'=G/\lambda $ and $G'=-\lambda G$ , together with a vertex  \textit{G' = 0  }located at  $G = 1$. 
For any value taken by the coupling ratio $\lambda$, the function $G'(\eta )$ reaches its extremes at the two roots of  $G(\eta )=e^{h} $. 
Also, as expected, in the case $\lambda =1$ Eq. \eqref{EQ_29_} identically reduces to \eqref{EQ_24_}. 
Being implicit, \eqref{EQ_29_} can generally not be solved for $G'$ as a function of $G$ by standard algebraic techniques. 
\newline

A practical yet accurate approximation for the function  $G'(G)$ predicated on Eq. \eqref{EQ_24_}, which removes the dependence on  $\eta$ in \eqref{EQ_19_b} and uncouples the system,  is proposed below.
\newline

For the positive branch  $G' \geqslant  0$ , for large $G$ the function $G'$ is asymptotic to $G'=G/\lambda $:  Eq. \eqref{EQ_29_} is thus reformulated as 
\begin{equation} \label{EQ_30_} 
\lambda \frac{G'}{G} =1-\frac{1}{G^{\lambda ^{2} +1} \left(1+\frac{1}{\lambda } \frac{G'}{G} \right)^{\lambda ^{2} } }  
\end{equation} 
Furthermore, the factor in parenthesis in the denominator always satisfies the following inequality 
\begin{equation} \label{EQ_31_} 
\left(1+\frac{1}{\lambda } \frac{G'}{G} \right)^{\lambda ^{2} } <e^{\lambda \frac{G'}{G} }  
\end{equation}

Upon approximating this factor by its exponential limit, Eq. \eqref{EQ_30_} becomes 
\begin{equation} \label{EQ_32_} 
e^{\lambda \frac{G'}{G} } \left( 1-\lambda \frac{G'}{G} \right)\cong \frac{1}{G^{\lambda ^{2} +1} }  
\end{equation}

Since the $G$-function is bounded by  $e^{h}$, the right hand side of  \eqref{EQ_32_} satisfies the following inequalities 
\begin{equation} \label{EQ_33_} 
e^{-h(\lambda ^{2} +1)} \le \frac{1}{G^{\lambda ^{2} +1} } \le 1 
\end{equation}

In order for \eqref{EQ_32_} to be consistent with  \eqref{EQ_33_}, the left hand side of  \eqref{EQ_32_} must at most be of order \textit{O}(1). Consequently, a Taylor expansion of the exponential function to  first order yields an explicit approximation for  $G'(G)$. For the positive branch  \textit{G' $\ge$ 0}  it is formulated as \eqref{EQ_34_a}; for the negative branch  $G' \le 0$,  $\lambda$\textit{-invariance} applied to \eqref{EQ_34_a} directly yields  \eqref{EQ_34_b}.
\begin{subequations}\label{EQ_34_}
    \begin{align}
       G'(G)&\cong \frac{G}{\lambda } \left( 1-\frac{1}{G^{\lambda ^{2} +1} } \right)^{1/2} \quad (\text{positive branch } G' \geqslant  0 )\label{EQ_34_a}\\
      G'(G)&\cong -\lambda G\left( 1-\frac{1}{G^{1/\lambda ^{2} +1} } \right)^{1/2}  \quad (\text{negative branch }  G' \le 0)  \label{EQ_34_b}
    \end{align}
\end{subequations}

Remarkably, the above approximate function $G'(G)$  satisfies the following three basic properties identical to those of an exact numerical solution of Eq. \eqref{EQ_29_}: 

\begin{enumerate}
\item  at its vertex, when $G = 1$, the function $G'(G)$ reaches $G' = 0$,

\item  for $G \gg  1$, as expected, the positive branch of the function $G'(G)$ is asymptotic to  $G'= G/\lambda$ whereas the negative branch is asymptotic to $G'= -\lambda G$,

\item  for $\lambda = 1$, the function $G'(G)$ reduces to the exact predicate expression  \eqref{EQ_24_}.

\end{enumerate}

Thus, in the $G-G'$ phase space, the explicit expressions \eqref{EQ_34_} represent approximate positive and negative branches of the ``skewed'' hyperbola defined by Eq. \eqref{EQ_29_} with the same orthogonal asymptotes. 
Upon comparing graphic representations of the explicit expressions \eqref{EQ_34_} to the exact numerical solution of \eqref{EQ_29_} for the implicit function $G'(G)$ it is found that the agreement is quite reasonable particularly for the positive $G'(G)$-branch when  $\lambda \geqslant 1$, and conversely for the negative branch when $\lambda \leqslant 1$. 
This is understandable in light of the above first two properties of \eqref{EQ_34_}. As $\lambda \to 1$  the approximation \eqref{EQ_34_} approaches the exact solution \eqref{EQ_24_}; for $\lambda \gg 1$  the graph of \eqref{EQ_34_}  exhibits two branches tightly bounded by their respective orthogonal asymptotes with the accuracy of this approximation  increasing with increasing $\lambda$. 
\newline

As intended, approximation \eqref{EQ_34_} effectively uncouples the system \eqref{EQ_19_} by explicitly removing the dependence on $\eta$ in the original ODE (18b): upon inserting the conservation Eq. \eqref{EQ_14_} into \eqref{EQ_34_}, Eq. \eqref{EQ_19_b} is replaced by a pair of two $\lambda$-\textit{invariant} \nth{1} order nonlinear ODEs for the hybrid species population $\xi (t)$ 
\begin{subequations}\label{EQ_35_}
    \begin{align}
       \dot{\xi }&=-\frac{h+1+\xi }{\lambda } \left( 1-\frac{e^{\xi (\lambda ^{2} +1)} }{(h+1+\xi )^{(\lambda ^{2} +1)} } \right) ^{1/2} \quad \text{  (positive }\eta \text{-branch: } \eta \geqslant  0) \label{EQ_35_a}\\
     \dot{\xi }&=\lambda (h+1+\xi ) \left( 1-\frac{e^{\xi (1/\lambda ^{2} +1)} }{(h+1+\xi )^{(1/\lambda ^{2} +1)} } \right)^{1/2} \quad \text{  (negative }\eta \text{-branch: } \eta \le 0)  \label{EQ_35_b}
    \end{align}
\end{subequations}

Evidently, for $\lambda = 1$ the two branches of \eqref{EQ_25_b}  are recovered. 
Even though $\xi (t)$ is not explicitly expressed as a function of time $t$, the arbitrary $\lambda \ne 1$ problem has thus been reduced to a pair of \textit{simple quadratures} for the function $t(\xi )$. 
As already stated, the function $\xi (t)$  oscillates between the $\lambda $-independent respective roots $\xi ^{-} (h)$ and  $\xi ^{+} (h)$ solutions of Eq. \eqref{EQ_18_}. 
The process for solving Eq. \eqref{EQ_35_} is identical to that of Eq. \eqref{EQ_25_b}: upon again choosing the time origin $t = 0$ when  $\xi _{0} =\xi ^{-} (h)$, a complete period is obtained by integration over the corresponding negative $\eta$\textit{-branch} in \eqref{EQ_35_b} until $\xi (t)$ reaches $\xi ^{+} (h)$, followed by an integration over the positive $\eta$\textit{-branch} \eqref{EQ_35_a} until $\xi ^{-} (h)$ is reached 
\begin{subequations}\label{EQ_36_}
    \begin{align}
      t(\xi )&=\int _{\xi ^{-} }^{\xi }\frac{1}{\lambda (h+1+x)}  \left( 1-\frac{e^{x(1/\lambda ^{2} +1)} }{(h+1+x)^{(1/\lambda ^{2} +1)} } \right)^{-1/2} dx \quad \text{  (negative }\eta \text{-branch}) \label{EQ_36_a}\\
     t(\xi )&=-\int _{\xi ^{+} }^{\xi }\frac{\lambda }{h+1+x}  \left( 1-\frac{e^{x(\lambda ^{2} +1)} }{(h+1+x)^{(\lambda ^{2} +1)} } \right)^{-1/2} dx   \quad \text{  (positive }\eta \text{-branch } )\label{EQ_36_b}
    \end{align}
\end{subequations}

The function $t(\xi )$ being monotonic and continuous on the respective integration intervals $\xi ^{-} \le \xi \le \xi ^{+} $ and $\xi ^{+} \geqslant  \xi \geqslant  \xi ^{-} $ its inverse function $\xi (t)$ exists and is unique, monotonic, and continuous on each interval. 
The LV problem is then completed for the function $\eta (t)$ by directly integrating the linear Eq. (18a) through standard numerical techniques. 
\newline

To assess the accuracy of the uncoupled approximate solutions \eqref{EQ_35_}, a comparison is made with the exact numerical solutions of the original coupled LV system \eqref{EQ_19_}. 
Upon using the respective values $\lambda = 2$ and $h = 2$ identical to those of Fig. 2 for the coupling ratio and system energy, Fig. 4 presents the comparison between the functions $\xi (t)$ and $\eta (t)$ respectively obtained by numerically integrating \eqref{EQ_35_} and  \eqref{EQ_19_} simultaneously through a standard 4th-order RK4 method.
From the figure it is observed that the ODEs \eqref{EQ_35_} provide a reasonably accurate solution for both functions $\xi (t)$ and $\eta (t)$ over an entire period, yet, when $\lambda > 1$, with an underestimation of the time taken to reach $\xi{}^{+}(h)$ compensated by an overestimation of the time to reach $\xi{}^{-}(h)$. As expected, the accuracy of the solutions obtained with approximations \eqref{EQ_35_} increases with increasing $\lambda$.

\begin{figure}[H]
  \centering
  \includegraphics[width=11cm]{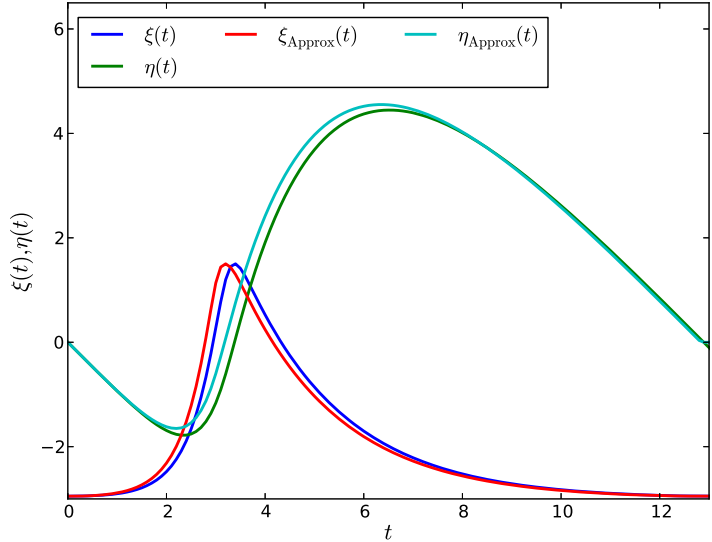}
  \caption{Solutions for $\xi (t)$ and  $\eta (t)$ as a function of time $t$  with $\lambda = 2$  and energy $h = 2$; comparison between RK4 numerical integration of  Eq. \eqref{EQ_19_} and Eq. \eqref{EQ_35_} }
  \label{fig:Fig5}
\end{figure}

From Fig. 4, regardless of the value of $\lambda$, the hybrid species population $\xi (t)$ is observed to oscillate with exponential-like growth and decay phases with its energy-dependent amplitude determined by the difference $\xi^{+}(h)$ - $\xi^{-}(h)$. 
\newline

Remarkably, in the high energy limit $(h \gg 1)$, upon keeping the leading asymptotic term in \eqref{EQ_35_}, the asymptotic behavior of the LV system becomes modeled as a system of two coupled \textit{ linear} \nth{1} order ODEs for each hybrid species.  In this asymptotic limit, together with the  linear ODE (18a) for $\eta (t)$,  the system admits trivial exponential solutions remarkably representative of the exact solutions of (18). For example, the asymptotic solutions $(h \gg 1)$ for the growth phase $(\xi ^{-} \le \xi \le \xi ^{+} )$ simply are
\begin{subequations} \label{EQ_37_}
    \begin{align}
    \xi(t)&=e^{\xi^{-}+\lambda t}-(h+1) \label{EQ_37_a}\\
    \eta(t)&=\frac{1}{\lambda}\left( \xi(t)-\xi^{-}(h) \right)-t  \label{EQ_37_b}
    \end{align}
\end{subequations}

The decay phase asymptotic solutions for $\xi(t)$ are obtained by $\lambda$-\textit{invariance}, namely $\lambda \to -1/\lambda$ together with $\xi^{-}(h) \to \xi^{+}(h)$.
\newline

Lastly, upon inserting the hybrid-species populations $\xi (t)$  and  $\eta (t)$ derived from Eqs. \eqref{EQ_36_} together with the transformation \eqref{EQ_11_} into the definition (5) of the prey and predator species, the respective standard solutions for the original populations $u(t)$ and $v(t)$ are fully recovered 
\begin{subequations}\label{EQ_37_}
    \begin{align}
      u(t)&=e^{\xi (t)-\lambda \eta (t)}  \quad \text{for preys } \label{EQ_37_a}\\
     v(t)&=e^{\xi (t)+\eta (t)/\lambda } \quad \text{for predators }\label{EQ_37_b}
    \end{align}
\end{subequations}

\section{Oscillation Period of the LV System}

The unique $\lambda$\textit{-invariance} property of  $\eta ^{\pm } (\xi ,\lambda )$ in \eqref{EQ_17_} directly enables to establish two fundamental properties of the LV system period. 
Consider the double mapping of Fig. 1 and follow in a counterclockwise direction the two branches $AB^{-}$ and $BA^{+}$ corresponding to the respective branches  $\eta ^{-} (\xi ,\lambda )$  and  $\eta ^{+} (\xi ,\lambda )$:  the negative branch $AB^{-}$  starts at  $\xi ^{-} (h)$ and ends at  $\xi ^{+} (h)$ and conversely for the positive $BA{}^{+}$ branch.  
Upon integrating \eqref{EQ_20_} over the $\xi$-variable and recalling the earlier definition $t=\sqrt{\alpha \delta } t'$ , the oscillation period  $T_{\lambda } (h)$  associated with the $\lambda$-mapping is directly obtained as a quadrature over these two branches \eqref{EQ_38_a};
here the negative sign for the second integral reflects integration from $\xi ^{+} $ to $\xi ^{-} $. Similarly for the $1/\lambda$-mapping the oscillation period is expressed as \eqref{EQ_38_b}
\begin{subequations}\label{EQ_38_}
    \begin{align}
      T_{\lambda } (h)&=\frac{1}{\sqrt{\alpha \delta } } \left( \int _{AB^{-} }\frac{d\eta ^{-} (\xi ,\lambda )}{h+\xi }  -\int _{BA^{+} }\frac{d\eta ^{+} (\xi ,\lambda )}{h+\xi }  \right) \label{EQ_38_a}\\
     T_{1/\lambda } (h)&=\frac{1}{\sqrt{\alpha \delta } } \left( \int _{AB^{-} }\frac{d\eta ^{-} (\xi ,1/\lambda )}{h+\xi }  -\int _{BA^{+} }\frac{d\eta ^{+} (\xi ,1/\lambda )}{h+\xi }  \right) \label{EQ_38_b}
    \end{align}
\end{subequations}

Upon recalling the $\lambda$\textit{-invariance}  property of Eq. \eqref{EQ_17_}, substitution into \eqref{EQ_38_b} establishes that:
\begin{equation} \label{EQ_39_} 
T_{\lambda } (h)=T_{1/\lambda } (h) 
\end{equation}

\begin{thm} 
For any value of the positive orbital energy  $h$ , the LV system oscillation periods respectively corresponding to the coupling ratio  $\lambda$  and its inverse $1/\lambda$  are equal. 
\end{thm}

Consequently, an exact, closed-form, regular expression for the nonlinear LV system oscillation period, valid for any value of the coupling ratio  $\lambda$  and any value of the orbital energy  $h $, is directly derived from \eqref{EQ_38_a} as an integral over the two branches of the  $\xi$ - $\eta$  mapping 
\begin{multline}
 \label{EQ_40_} 
T_{\lambda } (h)=\frac{1}{\sqrt{\alpha \delta } } \frac{\bigl( \eta ^{-} (-h,\lambda )-\eta ^{+} (-h,\lambda )\bigr) (\xi ^{+} -\xi ^{-} )}{(h+\xi ^{+} )(h+\xi ^{-} )} + \\ 
\frac{1}{\sqrt{\alpha \delta } }\int _{\xi ^{-} }^{\xi ^{+} }\frac{\eta ^{-} (x,\lambda )-\eta ^{-} (-h,\lambda )+\eta ^{+} (-h,\lambda )-\eta ^{+} (x,\lambda )}{(h+x)^{2} }  dx
\end{multline}

In Appendix 1, for any  $\xi \in \{{\xi }^-(h),\ {\xi }^+(h)\}$, the interval  ${\eta }^+\left(\xi ,\lambda \right)-{\eta }^-(\xi ,\lambda )$  is shown to be a positive increasing function of  $\lambda$ when $\lambda \geqslant  1$ (and decreasing when  0 $<$ $\lambda \le 1$) admitting respective lower and upper bounds, both of which are minimal when  $\lambda =1$. Together with Eq. \eqref{EQ_40_} this establishes:

\begin{thm}  
For any value of the positive orbital energy \textit{h} , the LV system oscillation period $T_{\lambda }(h)$ is an increasing function of  $\lambda$  for $\lambda \geqslant  1$ (decreasing for 0 $<$ $\lambda \le 1$) and the period is shortest for  $\lambda =1$.
\end{thm}

In the particular case when  $\lambda =1$, the exact LV system period $T_{1}(h)$  is uniquely expressed in terms of a universal energy function  $\Theta_{1} (h)$  as
\begin{equation} \label{EQ_41_} 
T_{1}(h)=\frac{2\pi }{\sqrt{\alpha \delta } } \Theta_{1} (h) 
\end{equation} 

The LV energy function $\Theta_{1} (h)$ introduced here is readily defined from \eqref{EQ_27_} as 
\begin{equation} \label{EQ_42_} 
\Theta_{1} (h)=\frac{1}{\pi } \int _{\xi ^{-} }^{\xi ^{+} }\frac{dx}{\sqrt{(h+1+x)^{2} - e^{2x}} }   
\end{equation} 

At small orbital energy ($h \ll 1$),  $\Theta_{1} (h)$ is directly expressed in terms of the complete elliptic integral of the first kind $\bm{K}(k)$ with its modulus $k$ 
\begin{equation} \label{EQ_46_}
\Theta_{1} (h) = \frac{1}{\sqrt{1+\sqrt{2h}}}\frac{2}{\pi} \bm{K}(k) \quad \text{ with } \quad   k = \sqrt{\frac{2\sqrt{2h}}{1+\sqrt{2h}}}
\end{equation}

A standard series expansion for $\bm{K}(k)$ yields
\begin{equation} \label{EQ_47_}
\Theta_{1} (h) = 1+\frac{1}{6} h + \frac {35}{432} h^2 + O(h^3)
\end{equation}

As expected, for small oscillation amplitudes, the integral (42) is  independent of the  energy $h$  and exactly equates $\pi$: hence  $\Theta_{1} (h)$ approaches unity in  (44) and the LV system period $T_{1}(h)$ is that of a harmonic oscillator with time factor $1/\sqrt{\alpha \delta}$, as already established  \cite{volterra1926}, \cite{waldvogel}.
\newline

At high orbital energy ($h \gg 1$), the contribution from the exponential term in \eqref{EQ_42_} is negligible over the integration interval except when $\xi$ approaches $\xi^{+} (h)$: since by definition $\xi \geqslant   \xi ^{-} (h)$, approximating the exponential term by its lowest value  $e^{2\xi ^{-} (h)} $ and performing the integration yields an asymptotic expression for $\Theta_{1} (h)$
\begin{equation} \label{EQ_43_} 
\Theta _{\text{asymp}} (h)\cong \frac{1}{\pi } \cosh ^{-1} \left( e^{\xi ^{+} (h)-\xi ^{-} (h)} \right)    \quad   \text{  with } h \gg 1
\end{equation}

When $\lambda \ne 1$ the exact LV oscillation period $ T_{\lambda } (h)$ is obtained by numerically solving the ODE system (18) as done for Fig. 2. Similar to Eq. (41), for each value of the coupling ratio $\lambda$, the period  $ T_{\lambda } (h)$ is then uniquely expressed in terms of universal LV energy functions  $\Theta_{\lambda} (h)$ 
\begin{equation} \label {EQ_50_}
T_{\lambda}(h)=\frac{2\pi }{\sqrt{\alpha \delta } } \Theta_{\lambda} (h) 
\end{equation}

As shown on Fig. 5 and consistent with Theorem 2,  for any value of the coupling ratio $\lambda$, each  function $\Theta_{\lambda} (h)$ is a monotonically increasing function of the system's energy $h$; so is the LV system period $ T_{\lambda } (h)$,  \cite{waldvogel}. Also displayed is the asymptotic approximation (45)  of the exact function  $\Theta_{1} (h)$  ; for $h \geqslant  3$ the difference between the exact solution and its asymptotic approximation is $\le 3\%$.

\begin{figure}[H]
  \centering
  \includegraphics[width=11cm]{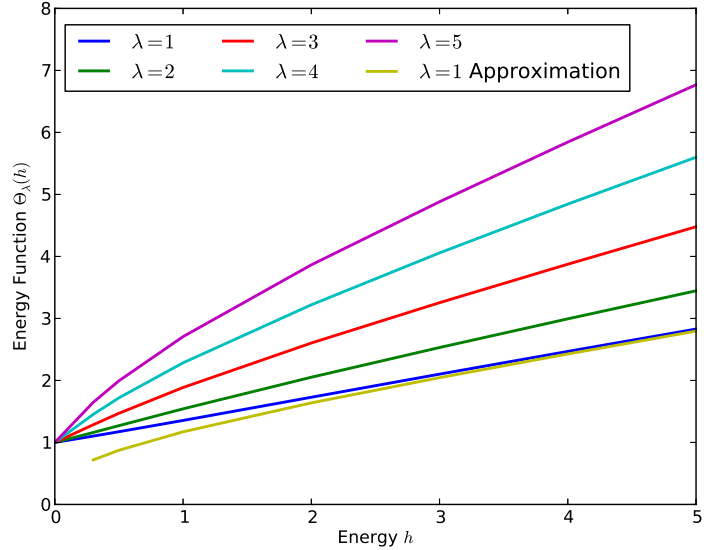}
  \caption{Energy function $\Theta_{\lambda}(h)$ for $\lambda = 1, 2, 3, 4, 5$ and asymptotic approximation for $\lambda =1$}
  \label{fig:Fig6}
\end{figure}

In this general $\lambda \ne 1$  case, an asymptotic formula for the LV system oscillation period  $T_{\lambda } (h)$ valid at high energy  $(h \gg 1)$ is  obtained from the asymptotic solutions (36). The  contribution  $T_{\lambda }^{+} (h)$ of the exponential growth phase of $\xi(t)$ to the period is readily obtained from Eq. (36b) since $\eta(t)=0$ when  $\xi(t)$ reaches its maximum $\xi^{+}(h)$; the  contribution $T_{\lambda }^{-} (h)$ of the decay phase is obtained by $\lambda$-\textit{invariance}. As a result the high energy $(h \gg 1)$ asymptotic expression for the LV system period $T_{\lambda } (h)$ simply becomes proportional to the sum of the $\xi(t)$-function growth and decay rates, $\lambda$ and $1/\lambda$, respectively

\begin{equation} \label{EQ_47_} 
T_{\lambda } (h)\cong \frac{\pi }{\sqrt{\alpha \delta } } \left(\lambda +\frac{1}{\lambda } \right)\left( \xi^{+} (h)-\xi^{-}(h)\right )
\end{equation}

This asymptotic formula which separately factorizes the LV system coupling from the $\lambda$-independent energy contribution satisfies both Theorem 1 and Theorem 2 since it is minimal when $\lambda = 1$. 
\newline

Shih performed an exhaustive review of  integral representations of the period of the two-species LV system: he compared the methods of Volterra \cite{volterra1926}, Hsu \cite{hsu}, Waldvogel \cite{waldvogel}, and Rothe \cite{rothe} and demonstrated that all of these representations are equivalent to his own solution in terms of a sum of convolution integrals \cite{shih1997}. 
Subsequent approximations of the LV system period in terms of power series \cite{shih2004} or perturbation expansions \cite{grozdanovski} have also been published. 
In Appendix 3, following the derivation of Rothe \cite{rothe},  we show that, even though not "planar'' in Rothe's sense (Eq. \eqref{EQ_8_}), the Hamiltonian \eqref{EQ_12_} based on hybrid-species populations provides a "state sum" $Z(\beta )$ identical to that of Rothe thereby establishing direct equivalence with Rothe's convolution integral for the LV oscillator period.

\section{Conclusion}

The coupled \nth{1} order non-linear ODE system for the LV problem of two interacting species has been re-formulated in terms of a single positive coupling parameter $\lambda$, ratio of the relative growth/decay rates of each species taken independently. 
Based on a Hamiltonian formulation combined with a linear transformation introducing "hybrid-species populations", a novel $\lambda$\textit{-invariant}  set of two \nth{1} order ODEs is obtained with one being \textit{linear}. 
As a result, an exact, closed-form quadrature solution of the LV problem is derived for any value of the coupling ratio $\lambda$  and any value of the system's energy (Eq. \eqref{EQ_21_}).

In the $\lambda = 1$ case, the LV problem completely uncouples and an exact explicit closed-form solution is expressed in terms of the orbital energy  $h$ as a simple quadrature for the population of one hybrid-species whereas the other hybrid species' solution is explicitly expressed in terms of the former.

In the $\lambda \ne 1$  case, a $\lambda$\textit{-invariant}  accurate practical approximation is derived that explicitly uncouples the LV system and provides a closed-form solution in terms of a single quadrature for one of the hybrid-species populations.
Remarkably, at high orbital energies $(h \gg 1)$, the original coupled non-linear LV ODE system totally uncouples and becomes entirely \textit{linear} admitting trivial asymptotic exponential solutions.

Further, as a consequence of $\lambda$-\textit{invariance}, for any value of the orbital energy $h$, the LV system oscillation period is shown to be identical when the coupling parameter $\lambda$ is inverted to  $1/\lambda$  and is smallest when  $\lambda =1$. 
In this particular case, an exact, closed-form expression for the non-linear LV system oscillation period is derived in terms of a universal LV energy function. In the $\lambda \ne 1$ case, a simple asymptotic expression for the LV system oscillation period is derived for high  energies $(h \gg 1)$.

\section*{Appendix 1}

This Appendix presents a proof of Theorem 2  introduced after Eq. \eqref{EQ_40_}. For the positive root  ${\eta }^+(\xi ,\lambda )$ , Eq. \eqref{EQ_14_} is written
\begin{equation*} \label{EQ_A1.1_} 
{\lambda }^2e^{\frac{\eta }{\lambda }}+e^{-\eta \lambda }=({\lambda }^2+1) U(\xi)  \tag{A1.1}
\end{equation*}

For any given value of $\xi \in \{{\xi }^-(h),\ {\xi }^+(h)\}$, since we seek a positive root and since by definition 
 $0\le e^{-\eta \lambda }\le 1$, this root admits a lower and an upper bound
\begin{equation*} \label{EQ_A1.2a_} 
  \lambda \ln \Bigg( \bigg(1+\frac{1}{{\lambda }^2}\bigg)U\left(\xi \right)-\frac{1}{{\lambda }^2}\Bigg)\ \le {\eta }^+(\xi ,\lambda )\le \lambda \ln \Bigg ( \left(1+\frac{1}{{\lambda }^2}\right)U\left(\xi \right) \Bigg)
\tag{A1.2a}
\end{equation*} 

Similarly, by $\lambda$-\textit{invariance}, the negative root satisfies
\begin{equation*} \label{EQ_A1.2b_} 
-\frac{1}{\lambda }\ln \Big(\left(1+{\lambda }^2\right)U\left(\xi \right)\Big)\le {\eta }^-\left(\xi ,\lambda \right)\le -\frac{1}{\lambda }\ln \Big( \left(1+{\lambda }^{2}\right)U\left(\xi \right)-{\lambda }^2\Big)
\tag{A1.2b}
\end{equation*}

From Eqs. (A1.2) the lower and upper bounding of the roots $\eta ^{\pm } (\xi ,\lambda )$ of Eq. \eqref{EQ_16_} enables to  prove Theorem 2. From Eq. \eqref{EQ_40_}, the period depends on the magnitude of the positive interval  ${\eta }^+\left(\xi ,\lambda \right)-{\eta }^-(\xi ,\lambda )$. Upon introducing the ``outer limit'' ${\mathrm{\Delta }}_{\mathrm{out}}\left(\xi ,\lambda \right)$  as 
\begin{equation*} \label{EQ_A1.3a_} 
{\mathrm{\Delta }}_{\mathrm{out}}\left(\xi ,\lambda \right)=\lambda \ln\Bigg(\left(1+\frac{1}{{\lambda }^2}\right)U\left(\xi \right)\Bigg)+\frac{1}{\lambda }\ln\Big(\left(1+{\lambda }^2\right)U\left(\xi \right)\Big)            \tag{A1.3a}
\end{equation*}

it is readily seen that  ${\mathrm{\Delta }}_{\mathrm{out}}\left(\xi ,\lambda \right)$ is a positive, increasing function of $\lambda$ when $\lambda \geqslant  1$ (and decreasing when  $\lambda \le 1$) whose partial derivative  $\partial {\mathrm{\Delta }}_{\mathrm{out}}(\xi ,\lambda )/\partial \lambda $  vanishes when  $\lambda =1$. Similarly, upon introducing the ``inner limit'' ${\mathrm{\Delta }}_{\mathrm{in}}\left(\xi ,\lambda \right)$  as
\begin{equation*} \label{EQ_A1.3b_} 
{\mathrm{\Delta }}_{\mathrm{in}}\left(\xi ,\lambda \right)=\lambda \ln\Bigg(\left(1+\frac{1}{{\lambda }^2}\right)U\left(\xi \right)-\frac{1}{{\lambda }^2}\Bigg)+\frac{1}{\lambda }\ln\Big(\left(1+{\lambda }^2\right)U\left(\xi \right)-{\lambda }^2\Big)          \tag{A1.3b}
\end{equation*}

it is also seen that  ${\mathrm{\Delta }}_{\mathrm{in}}\left(\xi ,\lambda \right)$ is a positive, increasing function of $\lambda$ when $\lambda \geqslant  1$ (and decreasing when  $\lambda \le 1$) whose partial derivative  $\partial {\mathrm{\Delta }}_{\mathrm{in}}(\xi ,\lambda )/\partial \lambda $  also vanishes when  $\lambda =1$.
Since the positive interval  ${\eta }^+\left(\xi ,\lambda \right)-{\eta }^-(\xi ,\lambda )$ obviously satisfies
\begin{equation*} \label{EQ_A1.4_} 
{\mathrm{\Delta }}_{\mathrm{in}}\left(\xi ,\lambda \right)\le {\eta }^+\left(\xi ,\lambda \right)-{\eta }^-(\xi ,\lambda )\le {\mathrm{\Delta }}_{\mathrm{out}}(\xi ,\lambda )              \tag{A1.4}
\end{equation*}

This proves Theorem 2.

\section*{Appendix 2}

 Upon recalling the definition \eqref{EQ_15_} of $U(\xi)$,  a series expansion for the quadrature solution \eqref{EQ_27_} is derived by first writing the integral as
\begin{equation*} \label{EQ_A2.1_} 
t(\xi )=\cosh ^{-1} \bigl(U(\xi )\bigr)+ \int _{\xi ^{-} }^{\xi } \frac{1}{\sqrt{1-U(x)^{-2} } }  dx               \tag{A2.1}
\end{equation*}

Since $1\le U(\xi )\le e^{h} $, a binomial expansion of the integrand with  binomial coefficients expressed in terms of the Gamma function $\Gamma (p)$ defined by its standard Euler integral of the second kind yields the solution in terms of a converging series 
\begin{equation*}  \label{EQ_A2.2_} 
t(\xi )=\cosh ^{-1} \bigl(U(\xi ) \bigr) + \sum _{p=0}^{\infty }\frac{\Gamma \left(\frac{1}{2} \right)}{\Gamma \left(\frac{1}{2} -p \right)\Gamma \left (p+1 \right)} \int _{\xi ^{-} }^{\xi } U(x)^{-2p} dx                \tag{A2.2}
\end{equation*} 

Each integral $I_{2p}(\xi)$ in the expansion \eqref{EQ_A2.2_} is of the form 
\begin{equation*} \label{EQ_A2.3_} 
I_{2p} (\xi )=\int _{\xi ^{-} }^{\xi }\frac{e^{2px} dx}{(h+1+x)^{2p} }                              \tag{A2.3}
\end{equation*}

Successive integrations by parts and substitution into \eqref{EQ_A2.2_} result in a slowly convergent series of exponential integral functions with positive argument of the form $\mathrm{Ei}\bigl({2p(h+1+\xi )}\bigr)$  where the integer $p$ is $1, 2, 3, \dots$.

\section*{Appendix  3}

Based on thermodynamics, Rothe \cite{rothe} established that the Laplace transform of the period function $T(h)$, in which $h$ is the system's energy, is the canonical state sum $Z(\beta )$ of the Hamiltonian \eqref{EQ_8_}, with $\beta \in (0,\infty )$ as the inverse of the absolute temperature, namely 
\begin{equation*} \label{EQ_A3.1_} 
Z(\beta )=\int _{-\infty }^{+\infty }\int _{-\infty }^{+\infty }e^{-\beta H(x,y)} dxdy=  \int _{0}^{\infty }e^{-\beta h}  T(h)dh              \tag{A3.1}
\end{equation*}

From Eqs. \eqref{EQ_9_} and \eqref{EQ_12_} together with the definition \eqref{EQ_13_} of the G-function, the LV system's Hamiltonian is 
\begin{equation*} \label{EQ_A3.2_} 
H(\eta ,\xi )=\left(\lambda+\frac{1}{\lambda}\right)\left(G_{\lambda} (\eta )e^{\xi}-\xi -1\right)              \tag{A3.2}
\end{equation*} 

For notation purposes, we introduce the reduced $g$-function $g_{\lambda}(\eta)$ defined as 
\begin{equation*} \label{EQ_A3.3_}
g_{\lambda}(\eta)= \lambda e^{\frac{\eta}{\lambda}}+{\frac{1}{\lambda}}e^{-\eta\lambda}		\tag{A3.3}
\end{equation*}

Consequently, upon inserting the Jacobian $\mid J \mid= \left(\lambda+\frac{1}{\lambda}\right)$ of the linear transformation \eqref{EQ_11_}
\begin{equation*} \label{EQ_A3.4_} 
Z(\beta )= \left(\lambda+\frac{1}{\lambda}\right)\int _{-\infty }^{+\infty }\int _{-\infty }^{+\infty }   e^{-\beta g_{\lambda}(\eta) e^{\xi }+\left(\lambda+\frac{1}{\lambda}\right)\beta (\xi +1)}d\xi d\eta            \tag{A3.4}
\end{equation*}

Upon substituting  $s=e^{\xi }$  with $s\in (0,\infty )$, \eqref{EQ_A3.4_} becomes 
\begin{equation*} \label{EQ_A3.5_} 
Z(\beta )=\left(\lambda+\frac{1}{\lambda}\right)e^{\beta\left(\lambda+\frac{1}{\lambda}\right) } \int _{-\infty}^{+\infty }\int _{0}^{\infty }s^{\beta\left(\lambda+\frac{1}{\lambda}\right) -1}   e^{-\beta s g_{\lambda} (\eta )}ds d\eta               \tag{A3.5}
\end{equation*} 

The integration over $s$ is expressed in terms of the Gamma function $\Gamma (s)$:
\begin{equation*} \label{EQ_A3.6_} 
Z(\beta )= \left(\lambda+\frac{1}{\lambda}\right)\left(\frac{e}{\beta}\right)^{\beta\left(\lambda+\frac{1}{\lambda}\right) } \Gamma\Bigl(\beta \left(\lambda+\frac{1}{\lambda}\right)\Bigr) \int _{-\infty}^{+\infty } (g_{\lambda}(\eta))^{-\beta \left(\lambda+\frac{1}{\lambda}\right)}d\eta               \tag{A3.6}
\end{equation*}

Together with the above definition of  $g_{\lambda}(\eta)$ this definite integral has been evaluated (see 3.314 in \cite{gradshteyn}); the $\lambda$-\textit{invariant} state sum $Z(\beta )$ thus becomes 
\begin{equation*} \label{EQ_A3.7_} 
Z(\beta )=\left(\frac{e}{\beta\lambda}\right)^{\beta\lambda } \Gamma(\beta\lambda ) \left(\frac{e\lambda}{\beta}\right)^{\left(\frac{\beta}{\lambda}\right)} \Gamma\left(\frac{\beta}{\lambda}\right)                    \tag{A3.7}
\end{equation*}

Although the Hamiltonian \eqref{EQ_A3.2_} is defined in the $\xi-\eta$ space, the result \eqref{EQ_A3.7_} for the state sum $Z(\beta )$ is identical to that of Rothe (Eqs. \eqref {EQ_9_} and \eqref{EQ_10_} in \cite{rothe}) who used the "planar" Hamiltonian \eqref{EQ_8_} in the $x-y$ space.  The derivation of the period then directly follows Rothe who defines a function $\tau (h)$ (Eqs. \eqref{EQ_15_}, \eqref{EQ_16_}, and \eqref{EQ_17_} in \cite{rothe})  whose Laplace transform is 
\begin{equation*} \label{EQ_A3.8_} 
\int _{0}^{\infty }e^{-\beta h}  \tau (h)dh=\left(\frac{e}{\beta }\right)^{\beta} \Gamma (\beta )                 \tag{A3.8}
\end{equation*}

Since our state sum  \eqref{EQ_A3.7_} is expressed as the product of two  Laplace transforms similar to \eqref{EQ_A3.8_}, use of the Hamiltonian \eqref{EQ_A3.2_} establishes that the period $T_{\lambda}(h)$ of the LV system \eqref{EQ_19_} is directly equivalent to that of Rothe. Upon recalling the earlier definition of time $t=\sqrt{\alpha \delta } t'$ , the period is formulated as a $\lambda$-\textit{invariant} convolution integral satisfying Theorem 1 with $\tau(h)$ defined above
\begin{equation*} \label{EQ_A3.9_} 
T_{\lambda}(h)= \frac{1}{\sqrt{\alpha \delta } } \int _{0}^{h}\tau \left(\frac{s}{\lambda}\right)\tau\bigl(\lambda (h-s)\bigr)ds                     \tag{A3.9}
\end{equation*}

\nocite{}

\bibliographystyle{abbrv}





\end{document}